# Moiré patterns in graphene - rhenium disulfide vertical heterostructures

Ryan Plumadore[1], Mohammed M. Al Ezzi[2,3], Shaffique Adam[2,3,4] Adina Luican-Mayer[1*]

[1]Department of Physics, University of Ottawa, Ottawa, Ontario, Canada
[2] Department of Physics, National University of Singapore, 2 Science Drive 3, 117551 Singapore
[3] Centre for Advanced 2D Materials, National University of Singapore, 117546, Singapore
[4] Yale-NUS College, 16 College Avenue West, 138527 Singapore

*luican-mayer@uottawa.ca

Vertical stacking of atomically thin materials offers a large platform for realizing novel properties enabled by proximity effects and moiré patterns. Here we focus on mechanically assembled heterostructures of graphene and $ReS_2$, a van der Waals layered semiconductor. Using scanning tunneling microscopy and spectroscopy (STM/STS) we image the sharp edge between the two materials as well as areas of overlap. Locally resolved topographic images revealed the presence of a striped superpattern originating in the interlayer interactions between graphene's hexagonal structure and the triclinic, low in-plane symmetry of $ReS_2$. We compare the results with a theoretical model that estimates the shape and angle dependence of the moiré pattern between graphene and $ReS_2$. These results shed light on the complex interface phenomena between van der Waals materials with different lattice symmetries.

## **Introduction**

Heterostructures of atomically thin two-dimensional materials offer a playground for realizing physical systems with properties that are finely tuned by interlayer interactions[1]. Examples include the presence of correlated states in moiré patterns of twisted graphene layers[2-7], observation of Hofstadter butterfly[8] in twisted graphene/ hBN layers or superconductivity induced in the edge states of a quantum spin Hall insulator[9]. Studies of graphene placed on semiconducting transitional metal dichalcogenides (TMDs)[10] revealed that such a substrate can provide improved screening and reduced scattering. Different types of periodic interlayer interactions resulting from heterostructures where graphene is layered with in-plane anisotropic TMDs have recently been explored using substrates such as like $ReSe_2$[11] or BP [12]. On BP it was reported to give rise to shear-strained superlattices resulting in pseudo-magnetic fields. For the case of graphene/$ReSe_2$ interfaces[11], twist-dependent moiré patterns were measured using scanning tunneling microscopy. These studies motivate further development and understanding of intricate moiré patterns formed at interfaces between crystals with distinct symmetries. Scanning tunneling microscopy and spectroscopy (STM/STS) are experimental techniques well-equipped to offer such insight, as they can visualize the structure and its heterogeneity at the atomic scale as well as give information about the electronic states of the system.

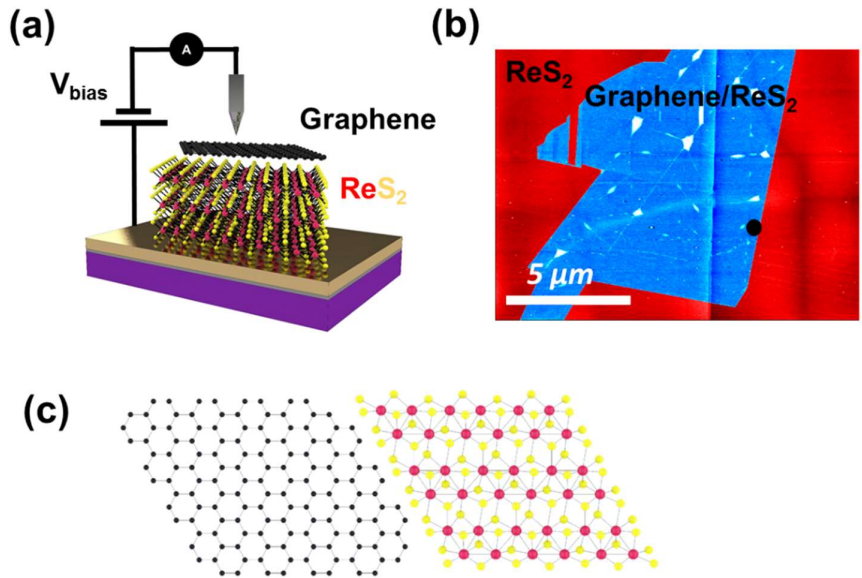

**Figure 1 (a) Schematic of the STM experiment on a ReS₂/Graphene heterostructure. (b) Atomic force micrograph of the area where graphene is stacked onto ReS₂. The black dot indicates the region where the data in Figure 2 was taken. (c) Graphene and ReS₂ lattices.**

Here we use room temperature, ultrahigh vacuum scanning tunneling microscopy and spectroscopy to locally resolve the properties of van der Waals heterostructures fabricated by vertical assembly of graphene and $ReS_2$, as illustrated in Figure 1(a). The heterostructures with atomically clean top surface were fabricated using a method where the entire heterostructure (here, graphene/$ReS_2$) is picked up by Polypropylene carbonate (PPC) polymer, then mechanically flipped upside down[9]. We placed this inverted structure onto a substrate which has been pre-patterned with gold contacts. Prior to measurement the samples are baked in forming gas (300°C for 10 hours), then annealed in vacuum (300°C for 10 hours). The atomic force microscopy topographic image in Figure 1(b) provides a closer view of the graphene/$ReS_2$ overlap area. We observe the presence of bubbles and ridges resulting from the transfer procedure, which serve as useful markers for positioning the STM tip during the measurement.

## Results and discussion

Topographic images acquired by STM identify the edge of the graphene on $ReS_2$ as shown in Figure 2(a). We measure a step height of 0.45 ± 0.01 nm, consistent with an atomically clean interface between the two layers. Also present in these topographic image are the bubbles and ridges formed during the fabrication steps. The typical height of the bubbles is 10-20nm and their typical width is 80nm. This aspect ratio is consistent with previous reports[13] and reflects the specific adhesion energy of the two layers. In Figure 2(b) we show the averaged scanning tunneling spectra over areas in Figure 2(a) both on the graphene/ReS₂ heterostructure as well as on the bare ReS₂. On the graphene, we observe the characteristic V-shaped spectrum[14], minimally doped away from the Dirac point as indicated in the left panel of Figure 2(b). This doping level is consistent to previously reported experiments where graphene was placed on TMDs like $MoS_2$ or $WSe_2$ [15]. On the ReS₂, the spectrum in the right panel of Figure 2(b) reveals the semiconducting gap, where we indicate the valence (VB) and conduction (CB) bands. The measured value of the gap is $\Delta E_{ReS2}$= 1.15 eV ± 0.2 eV, consistent with previous reports[16-20]. We note that in an STS measurement the Fermi level is at zero bias voltage, as specified in Figure 2(b). We find that in this system graphene's Dirac point is aligned with the conduction band edge of ReS$_2$. This is also consistent with the band alignment measured for MoSe$_2$ and WSe$_2$ [21].

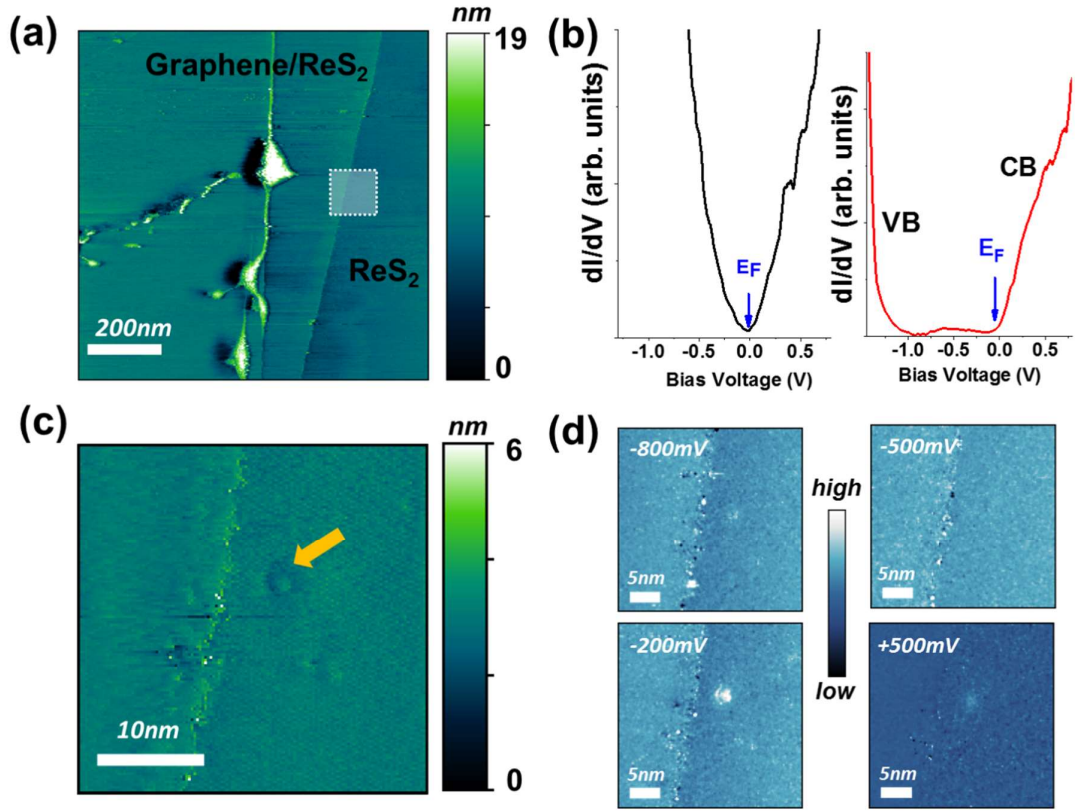

**Figure 2: (a) STM topographic image of an area at the boundary between the ReS2 covered with graphene and ReS₂. (b) Scanning tunneling spectroscopy on the graphene (left) and ReS₂ (right). (c) Topographic image at the boundary between graphene and ReS₂. (d) Conductance map in the same area as (c) taken at indicated bias voltage.**

Figure 2(c) shows a topographic image acquired in the dashed area of Figure 2(a) where the sharp edge of the graphene is visible[22]. On the ReS$_2$ we can resolve the atomic lattice, attesting to the surface cleanliness of the samples. At the same boundary, using differential conductivity maps, we spatially resolved the local density of states at the indicated bias voltages in Figure 2(d). The electronic contrast between graphene/ReS$_2$ and ReS$_2$ is most pronounced at -800mV, -500mV and -200mV, consistent with the spectroscopic data in Figure 2(b). We note that the feature marked by an arrow in Figure 2(c) represents a point defect in the ReS$_2$ crystal[19].

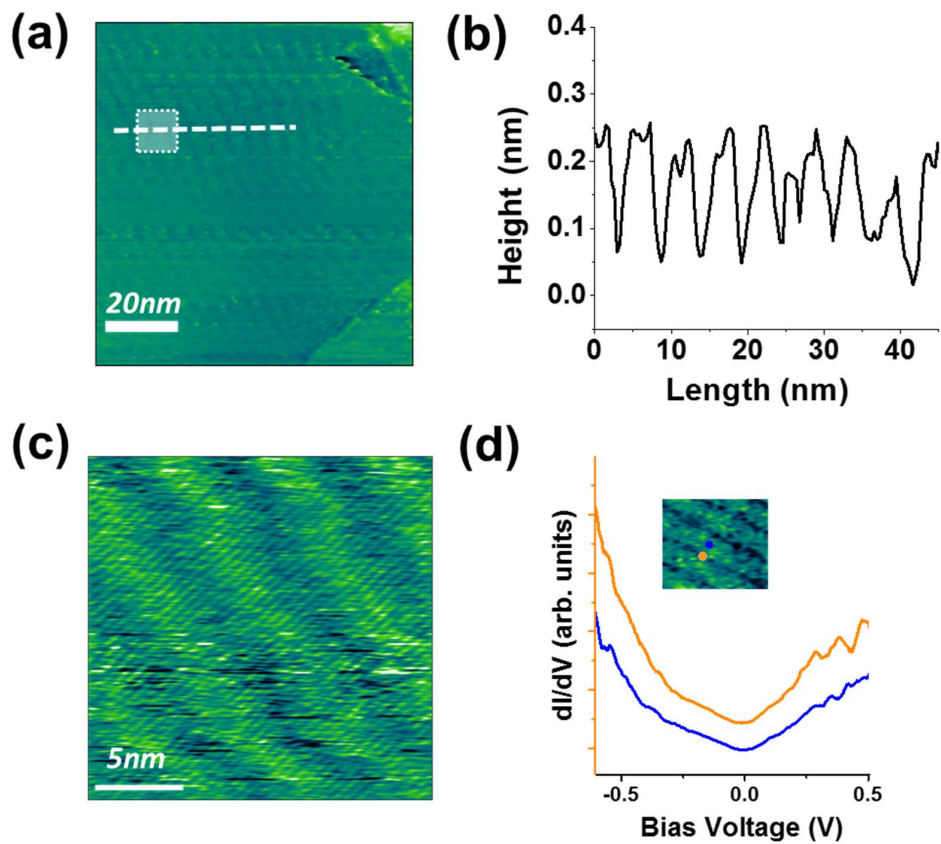

**Figure 3:** (a) Striped superpattern observed across the Graphene/ReS$_2$ interface in an STM topograph. (b) Height profile measured across the line indicated in (a). (c) Higher resolution STM topographic image corresponding to the area indicated in (a). (d) STS on the peaks and valleys of the striped pattern as indicated in the inset.

A characteristic topographic feature that the STM topographic images reveal in this system is presented in Figure 3(a), where we observe the appearance of a superpattern of parallel stripes. A height profile across the dashed line is plotted in Figure 3(b). The height corrugation of the pattern is approximately 200 pm, consistent across several areas measured on different samples and with different tips. The higher resolution topographic map shown in Figure 3(c) was taken across the area indicated in Figure 3(a) by a square; the atomic lattice of graphene as well as the striped pattern are visible. This superpattern of stripes resembles the features reported in both the graphene/BP and graphene/ReSe$_2$ systems [11-12]. Different from those systems, we note that the period of the superpattern we observe has been constant across different areas, samples, and tips. This suggests that either the structure we measured corresponds to an energy minimum for the twist at the graphene/ReS$_2$ interface, or that moiré patterns corresponding to other angles are less pronounced and therefore difficult to measure within the resolution of our STM. Indeed, theoretical calculations for the case of graphene/BP found that due to anisotropic orbital hybridization at the interface, the interlayer interaction could be orientation dependent[12]. Periodic superpatterns are sometimes associated with the presence of strain, which in graphene leads to the presence of pseudomagnetic field and associated quantized energy levels[12, 23-24]. In our experiments we have not observed signatures of such effects in the spectroscopic data at room temperature.

We propose that the experimentally observed one dimensional stripes are due to moiré patterns resulting from the interference pattern between graphene and ReS$_2$. Graphene has a hexagonal lattice with lattice parameters $a_1 = a_2 = 0.246$ nm, while ReS$_2$ has a triclinic lattice with $b_1 = 0.651$ nm $b_2 = 0.641$ nm and the angle between $b_1$ and $b_2$ is $118.9°$. Generally, for two materials with different crystal structures the moiré patters emerge due to the interference of extended unit cells of both materials and a possible relative twist angle between them[25]. For our case, the extended unit cells for graphene and ReS$_2$ should have primitive vectors given by $(m^*a_1, n^*a_2)$ and $(p^*b_1, q^*b_2)$ with a possible relative twist angle $\theta$ between them, where $m^*, n^*, p^*, q^*$ are four integers to be determined. To find the size of these effective unit cells and their relative twist angle we instead carry out our calculations in reciprocal space. The reciprocal lattice vectors for graphene are given by $G_{mn} = m\, G_1 + n\, G_2$, where $G_1$ and $G_2$ are the primitive reciprocal vectors for graphene. Similarly, the reciprocal lattice vectors for ReS$_2$ are $S_{pq}(\theta) = p\, S_1(\theta) + q\, S_2(\theta)$ where $S_1$ and $S_2$ are the primitive reciprocal vectors for ReS$_2$. The angle dependence is used to allow for a relative twist angle between the two materials. The real space wavelength characterizing the periodicity of the moiré patterns can be linked to reciprocal space quantities by

$$\lambda_{mnpq}(\theta) = |K_{mnpq}(\theta)|^{-1} \quad (1)$$

where $K_{mnpq}(\theta)$ is the difference between the wavevectors of both materials and is given by

$$K_{mnpq}(\theta) = G_{mn} - S_{pq}(\theta) \quad (2)$$

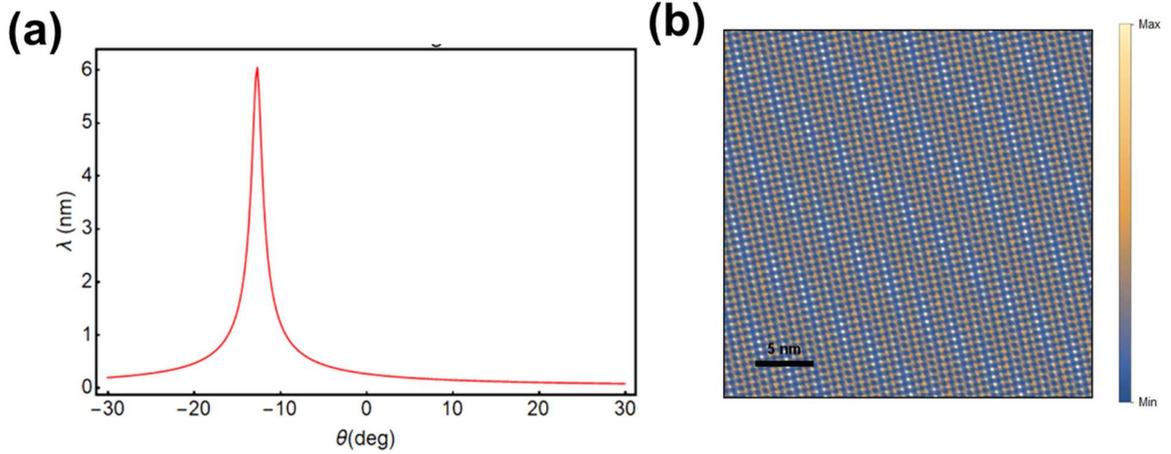

Figure 4: Moiré patterns. (a) Numerical calculation of moiré pattern wavelength as a function of twist angle. (b) Simulated real space image of the moiré pattern at a twist angle of 12 degrees with periodicity≈ $5.2\ nm$.

To find the possible moiré patterns we used the scheme proposed in [25] to utilize equation (1) and look for the patterns that resemble our experimentally observed patterns.

Our search in reciprocal space leads to the conclusion that the striped pattern is due to the extended unit cells with parameters $(m, n, p, q) = (3, 1, 6, 5)$. The moiré wavelength as a function of twist angle is shown in Figure 4(a). We can see the existence of a window of twist angles centered around 12 degrees where the wavelength is around 5 nm. Based on this calculation, we estimate the angle of rotation in the sample we measured to be about 12 degrees. We note that for most angles the superpattern is expected to have a period of below 1nm, which could be less pronounced in the STM topographic images. This is consistent with our observation of superpatterns that are larger (around 5nm) and only present in some areas of the sample. Figure 4(b) shows a simulated image for the moiré pattern in real space where the one-dimensional stripes can be seen with periodicity of approximately 5.2 nm.

## Conclusions

In summary, in this work we used scanning tunneling microscopy and spectroscopy to gain insight into the local electronic variations due to the interlayer interactions between graphene and ReS$_2$. We find a characteristic striped superpattern and discuss its origin as a moiré pattern between the two lattices. This works sheds light on the complex interfacial effects in novel heterostructures of two dimensional materials with different lattice symmetries.

## Acknowledgements

R.P. and A.L.-M. acknowledge funding from the National Sciences and Engineering Research Council (NSERC) Discovery Grant RGPIN-2016-06717. We also acknowledge the support of the Natural Sciences and Engineering Research Council of Canada (NSERC) through QC2DM Strategic Project STPGP 521420. S.A. acknowledges funding from Singapore Ministry of Education AcRF Tier 2 (MOE2017-T2-2-140).

## Data Availability Statement

The data that support the findings of this study are available from the corresponding author upon reasonable request.